\documentclass[12pt,a4paper,final]{iopart}

\usepackage{iopams}  
\usepackage{graphicx}
\usepackage[breaklinks=true,colorlinks=true,linkcolor=blue,urlcolor=blue,citecolor=blue]{hyperref}

\begin{document}

\title[Prime-Index Parametrization for Total Neutrino-Nucleon Cross Sections and {\it{pp}} Cross Sections \LaTeXe]
{Prime-Index Parametrization for Total Neutrino-Nucleon Cross Sections and {\it{pp}} Cross Sections}

\author{Ali R Fazely}
\address{Southern University, Baton Rouge, LA 70813, USA}
\ead{fazely@icecube.wisc.edu}

\begin{abstract}
A prime number  based parametrization for total neutrino-nucleon
cross section is presented. The method employs the relation between
prime numbers and their indices to reproduce neutrino cross sections
for neutrino energies from the $MeV$ to the $PeV$ regions where
experimental data are available. This prime-index relation provides
estimates of the neutrino-nucleon cross sections valid across many
decades of neutrino energy scales. The $PeV$ data are from the recently
published astrophysical $\nu_\mu + \bar \nu_\mu$ rates in the IceCube
detector as well as neutrino-nucleon cross section measurements.
A similar method has been employed for high energy $pp$ cross sections which
explains the $(\ln s)^{2}$ parametrization first proposed by Heisenberg.
\end{abstract}

\vspace{2pc}
\submitto{\JPG}

\section{Introduction}
Neutrino-nucleon and/or neutrino-nucleus cross section experiments
are divided into three categories; low, medium and high energies. The choice
of energy regions is motivated by neutrino production sources.
The low energy regime includes reactor, geoneutrinos and supernova
($SN$) neutrinos constituting neutrinos below $\approx 10\ MeV$ for reactor
and geoneutrinos and below $\approx 60\ MeV$ for $SN$ neutrinos. Above
$\approx 10\ MeV$ and below $\approx 350\ GeV$ neutrinos are produced in
accelerators, where the energy spectra and fluxes
as well as the flavor of neutrinos are known. In this medium energy region,
there is
also a contribution from the atmospheric neutrinos which extends to $100\ TeV$.
At energies above $100\ TeV$, neutrinos are produced predominantly by
astrophysical sources, recently observed by the IceCube
experiment.\cite{icecube}

Theoretical approaches to cross section calculations also follow
the above energy classifications, primarily because of the momentum
transfers
involved in the weak interaction processes.  At low energies, in the
$MeV$ region, the usual Shell Model, Random Phase Approximation
($RPA$), Quasiparticle Random Phase Approximation ($QRPA$) or
Effective Field Theory ($EFT$) calculations are usually
performed.\cite{petr} $RPA$, $QRPA$ and $EFT$ calculations
possess varying degrees of success depending on the target nucleus,
where the finer aspects of nuclear structure effects play a
significant role. In the medium energy region, Impulse Approximation
folded in with
a Fermi Gas Model, or Spectral Model calculations are often
used to reproduce the experimental data.\cite{petr} At higher neutrino
energies, above $0.1\ TeV$, Parton Model is mainly utilized
which we refer to as the $Standard\ Model\ (SM)$.\cite{reno} A
detailed review article of neutrino-nucleon/nucleus cross sections
across energy scales with description of various calculation methods
at different energies and comparison with experimental data was done
by Formaggio and Zeller.\cite{form}

The monotonically increasing behavior of neutrino cross section
measurements was the genesis for the low-energy approximation of Vogel
and Beacom.\cite{vb} This approximation includes first order corrections
in $\varepsilon=E_\nu/m_p$ and provides an estimate for $\bar\nu_e$ energies
below $E_{\bar\nu}<60\ MeV$.

\begin{equation}\label{eq:6}
\sigma\approx 9.53\, \times 10^{-44}\frac{p_e E_e}{\hbox{MeV}^2}\ \mbox{cm}^2,
\end{equation}
where $E_e=E_\nu \pm \Delta\hbox{ for }\bar{\nu}_e\hbox{ and }\nu_e$,
and $\Delta = M_n - M_p$ represent the $neutron-proton$ mass difference.

Strumia and Vissani\cite{stru} have derived the following approximation
shown in equation \ref{eq:7}, that agrees well with their full calculations,
below $E_{\nu}<300\ MeV$,
\begin{equation}\label{eq:7}
\sigma(\bar\nu_e p) \approx 10^{-43}\,\mbox{cm}^2~p_e E_e E_\nu^{-0.07056+0.02018\ln E_\nu-0.001953\ln^3 E_\nu},
\end{equation}
where energies are in $MeV$.

In the energy range of $GeV$ to $TeV$, a linear energy dependence for
neutrino and antineutrino nucleon cross sections of
$\sigma = 0.677E$ and $\sigma = 0.334E$ are derived from the measured
cross sections.\cite{pdg}

A rough estimate for $\nu N$ $Charged\ Current\ (CC)$ cross sections in ultra high energies
with $E > 10^6\ GeV$ and $E < 10^{12}\ GeV$, which is well above the $GZK$
limit, is provided by the following approximation\cite{gand},
\begin{equation}\label{eq:8}
\sigma(\bar\nu_e p)\approx 5.53 \times 10^{-36}\,\mbox{cm}^2~E_\nu^{0.363},
\end{equation}
where $E_\nu$ is in $GeV$.

\section{Primes and their Indices}
In this paper, we propose to estimate the total neutrino nucleon cross
sections by using the relation between prime numbers and their indices.
Prime numbers are a class of integers that are only divisible by themselves
and one. By this definition, the number one itself is not considered a
prime number. Some prime numbers and their indices are shown in
table \ref{tab:prim}.
\begin{table}[ht]
\caption{Some prime numbers with their positional indices}
\begin{center}
\begin{tabular}{llllll}
\hline\hline
Index & & & && Prime \\
\hline
1& &&&&2 \\
2& &&&&3 \\
.& &&&&. \\
8& &&&&19 \\
.& &&&&. \\
19& &&&&67 \\
.& &&&&. \\
114& &&&& 619 \\
.& &&&&. \\
619& &&&& 4567 \\
\hline\hline
\end{tabular}
\label{tab:prim}
\end{center}
\end{table}
Mathematically, it is a formidable task to determine if a number is prime.
Finding primes is usually done by the use of the $sieve\ theory$ which
is designed to count, or to estimate the size of sets of
integers. The $sieve\ theory$ also is used to sift out a set of prime
numbers up to some desired index. Note, we refer to the number in the
left column of
table \ref{tab:prim} as the index of a prime. The index of a prime is simply
the prime numbers ascending positional rank.
There is an asymptotic relation between primes and their indices. This is
due to $Gauss's$ prime number theorem that relates a prime to its index through
the following relation;
\begin{equation} \label{eq:9}
 \pi(p) \sim \frac{p}{\ln{p}}.
\end{equation}
Gauss later modified equation \ref{eq:9} to;
\begin{equation} \label{eq:10}
 \pi(p) \sim \int_2^p\frac{dx}{\ln{x}}.
\end{equation}
$Gauss's$ relation relates the index and its corresponding prime in an 
asymptotic
manner and the index and the prime approach the actual values as they tend to
infinity. Conversely, the prime number theorem is equivalent to stating
that the $i^{th}$ prime number is;

\begin{equation} \label{eq:12}
 p_{\pi(p)} \sim {\pi(p)}{\ln{\pi(p)}}.
\end{equation}

\section{Results and Discussions}
In this paper, we use relations between primes and their indices and
twin primes (a pair of primes which differ by 2, such as 11, 13 or 107, 109) 
and their indices to estimate total $CC$ and $Neutral\ Current\ (NC)$ 
neutrino-nucleon cross
sections and total $pp$ cross sections. Note, these estimates are empirically
driven and not to be considered as a replacement for the above-mentioned
physics-based approaches.

As mentioned above, we utilize the relation between a given prime and its
corresponding index to arrive at an estimate of the total cross section
for the $CC$ and $NC$ neutrino-nucleon scattering.
The method takes the index of the prime as the energy and the prime itself
then provides the cross section for that energy. We
assign the index to be in units of $MeV$. Note, the choice of $MeV$ is
made because it is of the order of the neutron-proton mass difference or of
the nuclear binding energy and is maintained regardless of the energy region.
The cross sections are then expressed in
units of $0.70$ of $10^{-42}\ cm^2$ or in units of $0.70\ ato-barn$
($1\ ab = 10^{-18}\ barn$) for $\nu_eN$ $CC$ cross sections.
For example, at a neutrino energy of $19\ MeV$ which is the index of the
prime number $67$, the cross section is $0.70 \times 67$ or $\approx 47\ ab$,
i.e. $\approx 47\times10^{-42}\ cm^2$.
A two-column look-up table such as table \ref{tab:prim} may be generated
where the first column is the index of the prime and it represents
the neutrino energy in $MeV$ and the second column is the prime
number multiplied by $0.70$. This represents the neutrino-nucleon cross
sections in $ab$. For antineutrinos, the second column is multiplied by
$0.26$ to produce the cross sections also in $ab$. Note, the $prime-index$
method provides a quick and accurate estimate of the neutrino-nucleon cross
sections
over many decades of energy scales and hence a reliable estimate for
interaction rates in various neutrino experiments.

Cross section estimates can also be obtained from equation \ref{eq:12}
where the index $i$ is the energy in $MeV$ and its corresponding prime, $p$,
represents the un-normalized cross sections in $ab$. This method introduces
a $22\%$ error at low energies and about $7\%$ error at high energies
when compared to actual primes and their indices.

\begin{figure}[ht]
\includegraphics[width=12.0cm]{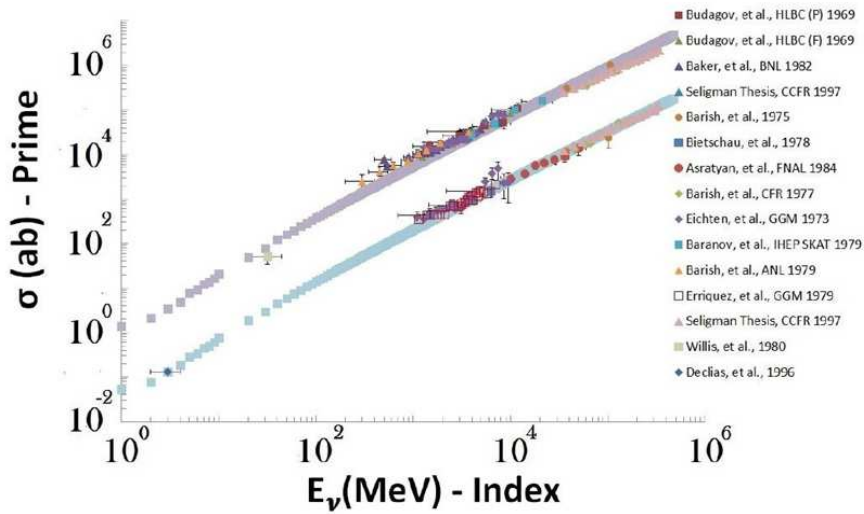}
\caption{The visible energy distribution for neutrinos and antineutrinos
on various nuclear targets. Note the data represents total neutrino and
antineutrino reaction cross sections with an isoscalar nucleon.
The antineutrino cross sections have been multiplied by $0.1$ for easier
visualization. The corresponding prime number distributions have been multiplied
by $0.70$ (blue bell squares) for neutrinos and by $0.26$ for antineutrinos
(dark sky blue squares).}
\label{fig:PI.eps}
\end{figure}

In figure
\ref{fig:PI.eps} the flux-averaged cross sections for many accelerator produced
neutrinos and antineutrinos reacting on various targets are shown.\cite{datum}
Note, the values of the antineutrino cross
sections data are multiplied by $0.1$ for ease of visualization.

\section{IceCube High Energy Astrophysical Neutrinos}
The IceCube collaboration published the observation of astrophysical
muon neutrinos in the IceCube Detector.\cite{sebas} The muons observed
are clear indications of $\nu_\mu$ and $\bar\nu_\mu$ $CC$
interactions in the IceCube detector. The rate for these muon neutrinos have
been compared
with the $SM$ calculations of Cooper-Sarkar, et al., \cite{sarkar}. The
$prime-index$ method
is also compared with those of reference \cite{sarkar}. Table \ref{tab:xsec}
shows these total $\nu_{\mu}\ CC$ cross sections $\geq\ 20\ TeV$ and their ratios. Note,
in obtaining the cross sections with the $prime-index$ method, the average of the
normalization for $\nu_\mu$ and $\bar\nu_\mu$ $CC$ cross sections of $0.70$
and $0.26$, i.e., $0.48$ was used.

\begin{table}[ht]
\caption{Comparison of $prime-index$ Cross Sections with those of reference \cite{sebas}.}
\begin{center}
\begin{tabular}{llll}
\hline\hline
Energy/GeV & $\sigma_T/pb\ (SM)$ & $\sigma_T/pb\ (prime-index)$ & Ratio \\
\hline
20000& 77& 177 & 2.30 \\
50000& 140& 469& 3.35 \\
100000& 210& 975& 4.64 \\
200000& 310& 2036& 6.57 \\
500000& 490& 5292& 10.80 \\
$1 \times 10^6$& 690& 10978& 15.91 \\
$2 \times 10^6$& 950& 22707& 23.90 \\
$5 \times 10^6$& 1400& 59078& 42.20 \\
$1 \times 10^7$& 1900& 121830& 64.12  \\
\hline\hline
\end{tabular}
\label{tab:xsec}
\end{center}
\end{table}

\begin{figure}[ht]
\includegraphics[width=12.0cm]{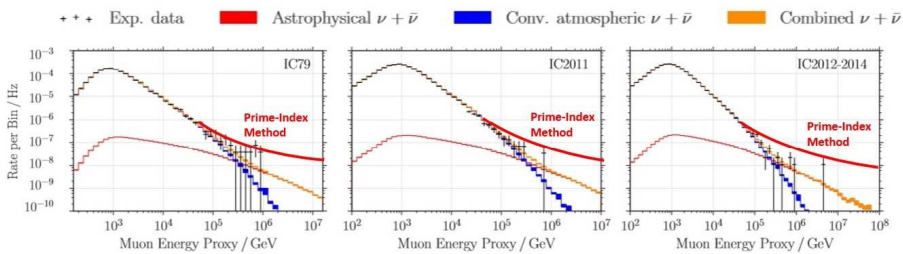}
\caption{The proposed $prime-index$ method superimposed on a figure showing
the experimental data from the published IceCube data of
reference \cite{sebas}.}
\label{fig:numu.eps}
\end{figure}

The $prime-index$ cross sections and the ratio to the $SM$ values were used
to calculate
the expected rate for $\nu_\mu + \bar\nu_\mu$ in the IceCube
detector. Figure \ref{fig:numu.eps} shows the $prime-index$ rate prediction.

The number of events observed above $0.5\ PeV$ where the
neutrinos are expected to be predominantly astrophysical are shown in table
\ref{tab:rate}. For comparison, the number of
events expected from the $prime-index$ method and the $SM$ are also listed.
Even with the limited number of events, there is strong
evidence that the $prime-index$ method is consistent with the observed number
of events in the IceCube detector.
\begin{table}[ht]
\caption{Comparison of IceCube data above $0.5\ PeV$ with $prime-index$ rate and with the $SM$}
\begin{center}
\begin{tabular}{llll}
\hline\hline
Energy/PeV & IceCube data& prime-index method & $SM$ \\
\hline
$> 0.5$ & $9\pm3$& $11\pm3.3$  & $1\pm1$\\
\hline\hline
\end{tabular}
\label{tab:rate}
\end{center}
\end{table}

\begin{figure}[ht]
\includegraphics[width=12.0cm]{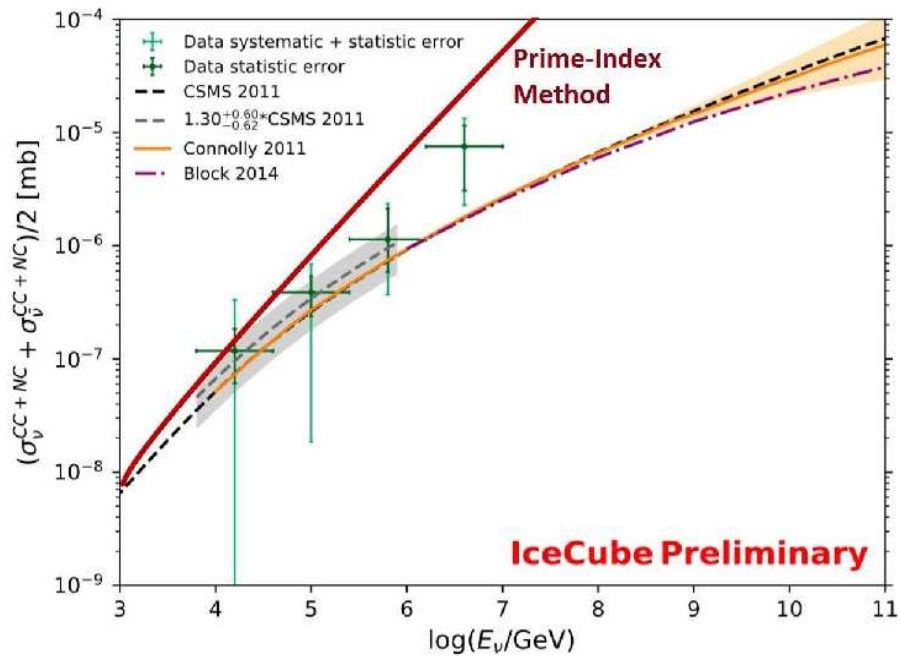}
\caption{The proposed $prime-index$ method superimposed on a figure showing
the $CC$ and $NC$ experimental data from the IceCube experiment. The prime-index
curve has a normalization factor of $0.69$.}
\label{fig:CCNC.eps}
\end{figure}

Recent IceCube analysis of neutrino and antineutrino cascade events for
$CC$ and $NC$ at high energies provide data to be compared with the
prime-index method. The ratio of the $CC$ cross section to the $CC\ + NC$ cross
sections can be written as;

\begin{equation} \label{eq:13}
\frac{\sigma_{CC}}{\sigma_{CC} + \sigma_{NC}} \approx{0.7}
\end{equation}
The normalization factors used in this paper for high energy 
astrophysical neutrinos are $0.48$ and $0.69$ for the data in 
figures \ref{fig:numu.eps}
and \ref{fig:CCNC.eps}, respectively. These factors  assume that the data
which were obtained with various neutrino and antineutrino flavors have 
roughly the
same flux. This is only true for more statistically significant
measurements. The abovesaid overall measurements are obtained with limited
statistics and each energy bin contain even less events. Hence, the use of
the above normalization factors should be only considered as a first step
approximation until further analyses of the IceCube data provide a larger
dataset.

At high neutrino energies, above $0.5\ PeV$, our proposed method when compared
to those obtained from the $SM$ begin to diverge by an order of magnitude and
increases as shown in table \ref{tab:xsec}. At $10^{12}\ GeV$, well
above the $GZK$ limit, cross sections obtained by the $prime-index$ method
are of the order $100\ mb$, $10^6$ times larger than those predicted by
$SM$. The proposed $10-km^3$ upgrade to the
IceCube detector, \cite{gen2} could provide the ideal laboratory for this
investigation. Also, further analysis of the IceCube data is highly
anticipated.
\section{Total {\it{pp}} Cross Section}
The total square of energy $s$ in the Center of Mass (CM) frame can be
written as;

\begin{equation}\label{eq:17}
s = 2{M_p}E + {M_p}^2
\end{equation}

\begin{figure}[ht]
\includegraphics[width=12.0cm]{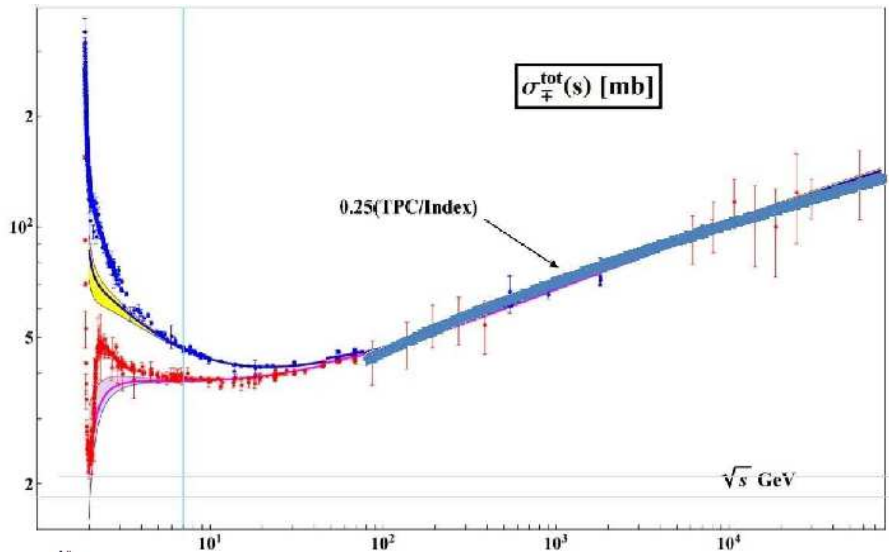}
\caption{The visible energy distribution for proton and antiprotons.
The total cross sections,i.e., the ratio of $TPC$ to index has been multiplied
by $0.25$ for normalizing to the experimental data.}
\label{fig:pp_xsections.eps}
\end{figure}

At high energies where $E\gg M_p$, we may neglect the term ${M_p}^2$;

\begin{equation}\label{eq:14}
\sqrt{s} = \sqrt{2M_pE}.
\end{equation}
The suggestion, originally by Heisenberg\cite{heis}, proposed a universal logarithmic
increase in the $pp$ cross sections of the form $(\ln s)^{2}$.
We note here that a $\ln^2$
relation implies a $twin\ prime\ -\ index$ quotient according to the
Brun theorem. Viggo Brun showed that the sum of reciprocals of the twin
primes was convergent.\cite{brun}
The $Brun's$ argument can be used to show that the number of twin
primes less than $N$ does not exceed $\frac{CN}{(\ln N)^{2}}$ and paved 
the way for the Hardy-Littlewood relation.\cite{hard}

\begin{equation}\label{eq:15}
\pi _{2}(x)\sim 2C_{2}{\frac {x}{(\ln x)^{2}}}
\end{equation}
In equation \ref{eq:15} $\pi_2$ is the index of a twin prime pair and $x$ is
a pair of twin primes. Hence, we can write the $(\ln x)^{2}$ as the ratio of
a twin prime to its index.
We generated the first $6.0\times 10^8$ twin primes by using their
$companion$ which we
refer to as the $Twin\ Prime\ Companion$ or $TPC$.\cite{faze, faze2}  
A $TPC$ is the composite sandwiched between a pair of twin primes.
As shown in figure \ref{fig:pp_xsections.eps}, the ratio of the $TPC$ 
to its index is the $pp$ cross section in $mb$ and the index is the energy in $GeV$.
Note, the ratio of $TPC$ to the index has been multiplied by $0.25$ to 
normalize it to the experimental data. 

\section{Summary}
In summary, the total neutrino and antineutrino cross sections
follow very closely the increase in the magnitude of primes
as a function of their positional indices. Neutrino and antineutrino
nucleon total $CC$ cross sections are estimated by the magnitude
of the prime numbers over six $(6)$ decades of neutrino energy where
experimental data are available. In this proposed method, the cross
sections are obtained from a two-column look-up
table where the first column is the index of the prime and it represents
the neutrino energy in $MeV$ and the second column is the prime
number multiplied by $0.70$. This represents the neutrino-nucleon cross
section in $ab$. For antineutrinos, the second column is multiplied by
$0.26$ to reproduce the cross sections in $ab$. At very high energies, the
$prime-index$ method tends to indicate increases in neutrino-nucleon cross
sections consistent with observed IceCube data.
Note, the $prime-index$ method also provides a quick and reliable rate
estimate for various neutrino experiments spanning $6$ decades of neutrino
energies. With the available published
astrophysical muon neutrinos the $prime-index$ method predicts
$11\pm3.3$ events vs. the observed $9\pm3$ events. Note, the $SM$
calculations predict $1\pm1$ event only. The method was also applied to
the IceCube cascade data which represents a mixture of $CC$ and $NC$.
A similar method, based on $twin\ primes$ and their indices, has been
employed for high energy $pp$ cross sections which explains the
$(\ln s)^{2}$ parametrization first proposed by $Heisenberg$.

\section*{Acknowledgments}
I am very grateful to Drs. Konstantin S. Kuzmin and Vadim A. Naumow for
promptly providing the total neutrino cross section data used in this paper.
I am grateful to Dr. G.P. Zeller and Professor Joseph A. Formaggio for
pointing to the whereabouts of more recent data. I am also very grateful to
Professor Francis Halzen for many wonderful discussions regarding this paper.

\end{document}